\documentclass{aa}
\usepackage{graphicx}
\usepackage{natbib}
\newcommand{\ROSAT}{{\it ROSAT}}
\begin{document}
   \title{Signs of WHIM in the soft X-ray background}

   \author{A. M. So\l tan \inst{1}, M. J. Freyberg \inst{2}
          \and
           G. Hasinger \inst{2}
          }

   \offprints{A. M. So\l tan}

   \institute{$^1$Nicolaus Copernicus Astronomical Center, Bartycka 18,
               00-716 Warsaw, Poland\\
              \email{soltan@camk.edu.pl} (AMS)\\
              $^2$ MPI f\"ur extraterrestrische Physik, Giessenbachstra{\ss}e,
               85748 Garching, Germany\\
              \email{mjf@mpe.mpg.de} (MJF)\\
              \email{ghasinger@mpe.mpg.de} (GH)
             }

   \date{Received 
}

  \abstract{
Small angular scale structure of the soft X-ray background correlated with 
the galaxy distribution is investigated. An extensive data sample from the \ROSAT\ and
XMM-Newton archives are used. Excess emission below $\sim 1$\,keV extending
up to at least $\sim 1.5$\,Mpc around galaxies is detected. The relative amplitude
of the excess emission in the $0.3-0.5$\,keV band amounts to $1.3\pm 0.2$\,\%
of the total background flux. A steep spectrum of the emission at higher energies
is indicated by a conspicuous decline of the signal above $1$\,keV.
The XMM-Newton EPIC/MOS data covering wider energy range than the \ROSAT\ PSPC
are consistent with a thermal bremsstrahlung spectrum with
${\rm k}T \la 0.5$\,keV.
This value is consistent with temperatures of the {\it Warm-Hot Intergalactic
Medium} derived by several groups from hydrodynamic simulations.
Correlation analysis allows for estimate of the average excess emission
associated with galaxies but the data are insufficient to constrain
physical parameters of the WHIM and to determine the
contribution of WHIM to the total baryonic mass density.
   }
   \authorrunning{A. M. So\l tan et al.\ }

\maketitle

\section{Warm-Hot Intergalactic Medium}

X-ray telescopes on board the satellites {\it EINSTEIN}, \ROSAT\, and {\it Chandra}
have firmly established the discrete nature of the extragalactic X-ray background
(XRB). High angular resolution provided by the imaging optics allowed for
isolation of various classes of point-like X-ray sources, leaving unresolved
only a small fraction of the total XRB flux. 

While the observational evidences are dominated by the discrete sources, theoretical
considerations indicate that in the soft X-rays still a non negligible fraction
of the XRB should remain in the form of the diffuse component. Apart from
clusters of galaxies which have been recognized sources of the extended
emission since the beginning of X-ray astronomy, a substantial
amount of plasma residing in the intergalactic space might be hot enough to emit
a noticeable amount of the X-ray flux.
Following \cite{cen99}, several groups performed hydrodynamical
simulations which have demonstrated that a significant fraction of baryons
in the Universe has not yet accumulated in stars and galaxies. It is estimated
that $30-40$\,\% of the baryonic matter fills the intergalactic space
(e.g. \citealt{dave}). This gas slowly falls toward potential wells created by the
(non-baryonic) dark matter. The density and temperature of the infalling material
increase and around mass concentrations arise halos of the hot plasma. 
The spatial distribution and physical parameters of the diffuse gas in the
local Universe ($z\la 0.4$) are distinctly different from both the matter
in galaxies and plasma in clusters of galaxies. This constituent of the
baryonic matter is known as WHIM -- {\it Warm-Hot Intergalactic Medium}
(\citealt{dave}).

The quantitative characteristics of the WHIM as a function of redshift depend
on the rate the gas is accumulating. The process is determined by the evolution
of the gravitational potential and by the non-gravitational heating, so called feedback
(e.g. \citealt{zhang}). Details of the feedback, i.e. transfer of energy and
matter from galaxies back to the intergalactic medium have been investigated
extensively in the literature (e.g. \citealt{dave}, \citealt{bryan},
\citealt{croft}). Although all these theoretical studies ascertain
the WHIM contribution to the soft XRB, the quantitative estimates of the WHIM
emission are still rather uncertain.

A potential contribution to the soft XRB from the diffuse component is strongly
constrained by the source counts. It appears that the integrated flux produced
by sources above the {\it Chandra} limit amounts to $94.3^{+7.0}_{-6.7}$\,\%
in the $0.5-2.0$\,keV band (\citealt{moretti}). A smooth extrapolation of counts
down to sources fainter by an order of magnitude than the present-day limit
gives $96$\,\% of the XRB. Consequently, the diffuse fraction cannot exceed
$4$\,\%, though this estimate is subject to large uncertainties.
Deep counts in the Lockman Hole by
\cite{worsley} using the XMM EPIC cameras show that sources above the
detection threshold of few $\times 10^{-16}$erg\,cm$^{-2}$s$^{-1}$ generate
in the $0.5-2.0$\,keV band $\sim 90$\,\% of the integral XRB. The resolved fraction
below $0.5$\,keV  is consistent with $100$\,\%, although, with large uncertainty
due to lack of precise measurement of the total flux.
Moreover, \cite{soltan03} estimated that roughly $1$\,\% of the soft XRB
constitutes purely diffuse component (unrelated to the WHIM emission) due to
Thomson scattering of X-ray photons.
Thus, the upper limits established for the WHIM are rather restictive and
impose important constraints on the feedback models (\citealt{bryan}).

Despite the poor understanding of processes involved and limited precision
of calculations, hydrodynamical simulations provide information on
physical parameters and spatial structure of the WHIM. The advanced models
are flexible enough to conform to the observational limits. For instance,
the simulations
by \cite{croft} predict that in the $0.5-2.0$\,keV energy band $\sim 6$\,\%
of the XRB is generated by the intergalactic gas with temperatures in
the range of $10^5 - 10^7$\,K.

The low emission strength makes the direct detection of the WHIM signal extremely
difficult. To search for traces of the extragalactic thermal radiation,
\cite{kuntz} analyzed the surface brightness distribution of the
\ROSAT\ All-Sky Survey (RASS) at high galactic latitudes. They found that
after subtraction of galactic and extragalactic sources and the local extended
emission, the data are consistent with the smooth thermal component with
${\rm k} T = 0.23$\,keV and amplitude of
$1.20\times 10^{-8}$\,erg/(s\,cm$^2$\,sr\,keV) in the ${3\over 4}$\,keV
band. In the discussion the authors stated that the observed signal
is probably a mixture of the WHIM and the Galactic halo emission.

All the simulations show that the WHIM emission is strongly correlated
with the galaxy distribution. Thus, the straightforward way to look for
the WHIM presence is to measure the XRB flux around galaxies and clusters.
Recently we have investigated the cross-correlations between the soft X-ray
maps of RASS and the galaxy distribution using the Lick counts
(\citealt{shane}) and Abell cluster catalog (\citealt{abell58},
\citealt{abell89}). Coarse pixel binning of the RASS ($12\arcmin \times
12\arcmin$) and Lick counts ($10\arcmin\times 10\arcmin$) allowed only
for the analysis of the correlation signal at relatively large angular
separations, effectively above $0\fdg 3$ (\citealt{soltan02}; references
to earlier work on this subject therein). Our investigation was also
impeded by the narrow useful energy range and the low energy resolution of
the RASS maps. The data we used were binned in three overlapping energy
bands: R5, R6 and R7 centered at $0.8$, $1.1$, and  $1.5$\,keV, respectively
(\citealt{snowden94}).

Our calculations revealed a clear excess of the soft emission associated with
the galaxies and clusters. The effect was measured at separations 
$0\fdg 3 - 2\fdg 1$ for the Abell ${\rm DC}\le 5$ clusters and
up to $3\fdg 1$ for the Lick counts. However, the signal amplitude
was weak and we were not able to put strong constraints on the temperature
of the emission. At a confidence level of $90$\,\% the temperature
${\rm k}T$ was below $1.0$\,keV (assuming the thermal
bremsstrahlung mechanism).

According to simulations, the surface brightness of the WHIM emission increases
sharply as the distance to the gravitational center gets smaller.
In the present investigation we analyze the distribution of the XRB intensity
around galaxies at few arc min scales, where the expected signal should
by stronger. In Sect.~\ref{data} the observational material is described.
Calculations and main results are presented in Sect.~\ref{signal} while
in Sect.~\ref{future} we shortly discuss limitations of the present work.

\section{Observational data \label{data}}

Both simulations (\citealt{croft}) and extrapolation of our results based
on the RASS (\citealt{soltan02}) show that the strength of the WHIM emission
is low in comparison with the total soft XRB even within $10$ arc min
from a typical galaxy in the sample. It appears that the
amplitude of the the cross-correlation function (CCF) of the soft XRB
and galaxy distribution generated by WHIM at separations above
$2\arcmin - 4\arcmin$ would not exceed a couple of percent. To detect
fluctuations of the XRB at such low level, scrupulous analysis of
an extensive data sample is required.

\subsection{PSPC data}

The central area of the \ROSAT\ PSPC field of view (unobstructed by the window
supporting structure) is a circle of $\sim\!27\arcmin$ diameter. Similar
size, albeit of more complex shape, have the field of view of the XMM-Newton
EPIC/MOS cameras. The excellent characteristics of the \ROSAT\ X-ray telescope
-- PSPC combination make this instrument well suited for searches of
the low amplitude fluctuations of the XRB. The PSPC has strikingly smooth
sensitivity over the entire field of view. The X-ray telescope introduces
practically negligible vignetting within the central area and contamination
of the accepted counts by the particle background is also quite low.

In a single \ROSAT\ observation with the exposure time of $10\,000$\,s,
typically $400$ PSPC counts are recorded in the central area in the band R5
and R6, and $\sim 300$ counts in R7 (if there are no bright sources in
the field). To detect relative fluctuations with the amplitude of
$\sim 0.01$ at high significance level, one needs a large number of
observations. In order to maximize the signal-to-noise ratio, all the
\ROSAT\ PSPC pointings at high galactic latitude, devoid of bright
or known extended sources, available in the archive have been used.

Since we concentrate on the diffuse extragalactic signal, only
pointings satisfying stringent criteria have been included in the analysis.
In particular, pointings at known extended sources (SN remnants, clusters
of galaxies, nearby galaxies) have been excluded. In selecting ``good''
observations we generally followed criteria used by \cite{soltan01} to build
a sample of observations suitable for the investigation of the autocorrelation
function of the soft XRB. In the present analysis we used also observations
at high galactic latitudes from the south galactic hemisphere.

From nearly 300 PSPC observations we have selected 217 for further
processing. The sample contained pointings with the Galactic hydrogen
column density below $7\times 10^{20}$\,cm$^{-2}$. Next, all the
sources detected in each energy band have been removed. We have
applied the Poissonian statistics to search for point-like enhancements
exceeding the $3\sigma$ threshold. Such procedure effectively eliminates
all high amplitude non-homogeneities of the count distribution and
substantially reduces noise of the CCF. Also, circular areas around all
clusters and groups of galaxies found in the NASA/IPAC Extragalactic
Database (NED) have not been used in the CCF calculations.
After the removal of all the sources and clusters, 
total numbers of counts in the data amount to $119\,000$ in the band R5,
$102\,000$ in R6 and $69\,000$ in R7.

\subsection{EPIC/MOS data}

A large collective area of the X-ray telescopes of the XMM-Newton Observatory
gives higher count rates as compared to the \ROSAT\ PSPC. Also a wide energy
range and good energy resolution contribute to the unique quality of the
data. However, investigation of the extended diffuse emission is hampered by
strong vignetting of the telescope, highly variable non--X-ray background
and imperfections of the CCD detectors. All these effects could be accounted
for with relative ease in case of the point source analysis, but pose
a serious problem for the investigation of the smoothly varying signal.
Elaborated procedures to deal successfully with the instrumental effects
in the context of the background measurements are presented by \cite{lumb},
\cite{read} and \cite{deluca}. These investigations allow to measure
the absolute level of the cosmic signal with the accuracy of several
percent. In the present investigation we do not attempt to determine
the net XRB flux, but we concentrate on the XRB fluctuations correlated
with the galaxy distribution. Although the expected amplitude of this signal
is smaller than uncertainties of the total XRB measurements, the correlation
analysis can be used to isolate effectively the WHIM emission.

To select observations suitable for the present investigation, criteria
analogous to the PSPC pointings have been applied. In preparation of
the data we generally followed prescriptions by \cite{snowden02}.
The data have been
divided into four energy bands: $0.3 - 0.5$\,keV (band X1), $0.5 - 1.0$\,keV
(X2), $1.0 - 2.0$\,keV (X3) and $2.0 - 4,5$\,keV (X4). For the band X1
only pointings with the Galactic hydrogen column density
$N_{\rm H} < 5\times 10^{20}$\,cm$^{-2}$ have been used, for band X2 --
pointings with $N_{\rm H} < 1.5\times 10^{21}$\,cm$^{-2}$, and for bands X3
and X4 -- pointings with $N_{\rm H} < 2\times 10^{21}$\,cm$^{-2}$. All the
data in the public archive obtained with the MOS1 and MOS2 detectors have
been used. After excluding pointings with bright sources and known extended
sources, the total number of usable pointings reached 150. All point-like
enhancements present in the data have been removed in a similar way as
in the PSPC observations.

Because of the vignetting the count distribution within the field of view
is highly nonuniform. Exposure maps created for each energy band include
vignetting effects. To compensate vignetting, one should correct
the image focused by the X-ray telescope using the exposure map.
However, amount of contamination by the non-vignetted
counts varies in time and for each pointing is different
(\citealt{read}). In effect, the vignetting corrections
are not adequately described by the exposure map and have to
be determined for each observation separately. Also, the source extraction
introduces some additional bias for the vignetting amplitude. Amount of
the removed photons depends on the exposure time (for longer exposures
a larger fraction of the XRB is decomposed into recognized sources). Thus,
the deep exposures are relatively more contaminated by the non-vignetted
counts.

To minimize gradients of the count distribution generated by the telescope
the following procedure has been applied.
We assumed that some, a priori unknown, fraction of counts, $f$, is subject
to the vignetting while the remaining $1-f$ counts are free from the vignetting
and are distributed randomly in the field of view. 
Thus, the large scale variations observed in the integral counts have smaller
relative amplitude than implicated by the exposure map, $EM$. The effective
exposure correction, $EC$, applicable to the actual observation
was assumed to have the form:
\begin{equation}
EC = (1-f)\cdot t_{\rm av} + f\cdot EM\,,
\end{equation}
where $t_{av}$ is the average exposure time.
The observed counts were divided by the $EC$ array and the resulting
``corrected'' counts were fitted to the flat distribution. The value of $f$
was adjusted using the maximum-likelihood method.
The calculations have been performed separately for each observation,
energy band (X1, ..., X4) and detector (MOS1 and MOS2)

\subsection{Galaxy data}

For all the X-ray observations (PSPC and EPIC/MOS) the
NED has been searched for galaxies
brighter than $20$ mag or within the redshift range of $0.02 - 0.20$.
Nearly $3400$ objects have been extracted from the database.
Clearly, the selected galaxies do not constitute a complete or
homogeneous sample. Pointings are covered by a variety of galaxy
surveys. In effect, the number of galaxies within the field of view of
the individual pointing depends on the specific survey characteristics
and to less extent reflects statistical properties of the
galaxy population at the selected magnitude limits.
Nevertheless, galaxies found within the field of view of
each pointing separately represent an unbiased set of objects suitable
for calculations of the CCF with the X-ray photon distribution.

\section{Diffuse emission in the PSPC and EPIC/MOS observations \label{signal}}

To estimate the average intensity of the XRB at distance $\theta$ from
a randomly chosen galaxy  we use the formula:
\begin{equation}
\rho(\theta) = {\sum n_{\rm cnt} \over \sum n_{\rm pxl} \cdot t_{\rm exp}}\,,
\label{ccf_def}
\end{equation}
where the sums extend over all pointings and all galaxies, $n_{\rm cnt}$
denotes total number of counts recorded in $n_{\rm pxl}$ pixels separated
by angle $\theta$ from the galaxy and $t_{\rm exp}$ is the exposure time
of the observation. All the PSPC data were binned into $8\arcsec \times 8\arcsec$
pixels and EPIC/MOS into $4\arcsec \times 4\arcsec$. In the absence of any signal
correlated with galaxies, one should expect $\rho(\theta) = \overline {\rho}$,
the average XRB flux.

\subsection{PSPC}

\begin{figure}
\resizebox{\hsize}{!}{\includegraphics[width=0.8\linewidth]{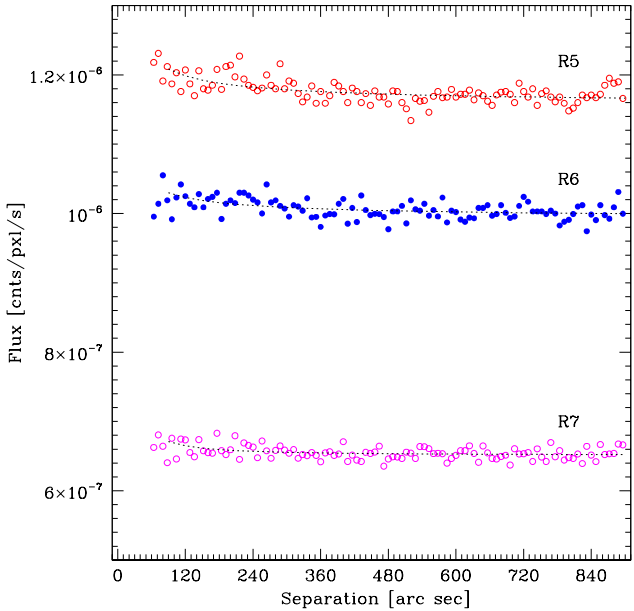}}
\caption{The countrate distribution in three \ROSAT\ PSPC
energy bands vs. angular distance to the galaxy averaged over the
galaxy sample. Power law fits are shown with dashed lines.
\label{pspc_ccf_1}
}
\end{figure}

\begin{figure}
\resizebox{\hsize}{!}{\includegraphics[width=0.8\linewidth]{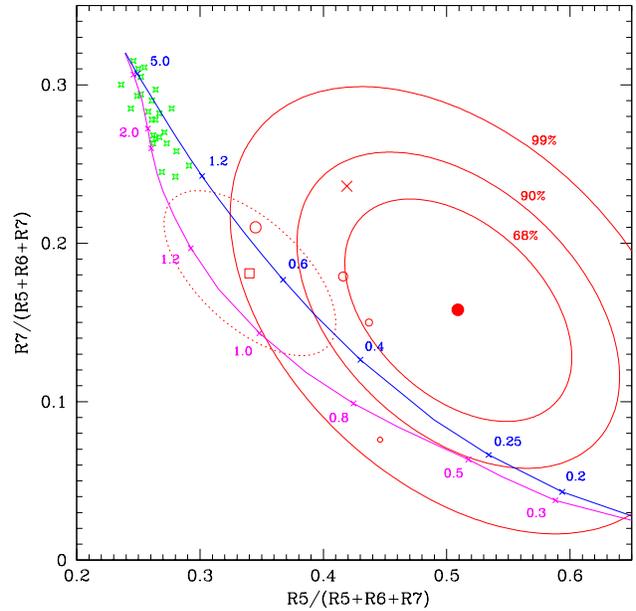}}
\caption{The X-ray ``colour-colour'' diagram for the PSPC observations.
The excess emission surrounding galaxies is shown with the full dot;
solid ellipses indicate $68$\,\%, $90$\,\% and $99$\,\% confidence limits.
The cross above the center of the plot indicates the colours of the average
total XRB.
The thermal bremsstrahlung emission (upper curve) and the plasma emission with
metal abundances at $20$\,\% of the cosmic abundances (lower curve) are shown;
labels denote temperature in keV. Colours of the enhancements found
in the RASS maps are shown with the open symbols. Square and circles
indicate emission of halos around Abell clusters (with the $90$\,\%
confidence region defined by the dotted ellipse) and galaxies in the Lick
counts, respectively (see the text for details).
Stars at the upper left show $25$ bright
\ROSAT\ clusters of galaxies.}
\label{r5_r7}
\end{figure}

The distributions of $\rho(\theta)$ in the three energy bands, R5, R6, and R7,
are shown in Fig.~\ref{pspc_ccf_1}. Scatter of points for each energy band
is slightly larger than the statistical noise 
implied by the number of ``galaxy -- X-ray count'' pairs. Additional
fluctuations of the $\rho(\theta)$ are generated by weak sources not removed
from the data. A quantitative analysis of plots in Fig.~\ref{pspc_ccf_1}
reveals weak but statistically significant systematic trends in all three bands.
The null hypothesis that $\rho(\theta) = const.$ over the separation
range of $2\arcmin - 15\arcmin$ is rejected at $5.4\sigma$ in the band
R5, $4.5\sigma$ in R6, and $2.9\sigma$ in R7. The S/N ratio is assessed
under the assumption that uncertainties of all points (for each band
separately) are equal. To test the efficiency of our analysis,
analogous calculations have been performed in which we searched for the
possible enhancements of the soft X-ray emission around quasars and known weak 
X-rays sources. Since the great majority of objects in those two categories are
much more distant than objects in the galaxy sample, we have not expected
to detect any correlation signal. Unfortunately, the number of objects in this
``test sample'' was much smaller than number of galaxies. Because of that
uncertainties of the CCF estimates have been substantially larger.
Nevertheless, the differences between the galaxy and the test sample
were significant. No X-ray enhancements around the test objects have been
detected and in the soft band R5 the test data were incompatible with
the galaxy signal at the level of $3.9\sigma$.

To measure the excess emission correlated with the galaxy distribution linear
fits to the plots in Fig.~\ref{pspc_ccf_1} have been calculated. The slopes
of those fits for three energy bands determine the enhancements potentially
generated by the WHIM.
These data were then used to construct the X-ray ``colour-colour''
diagram shown in Fig.~\ref{r5_r7} in the same way as the diagrams for the RASS
maps in \cite{soltan02}. A large solid dot indicates a position of the excess
emission. Solid ellipses define $68$\,\%, $90$\,\% and $99$\,\% confidence limits
based on the simulations. The large cross situated above the center of the
plot gives the average colours of the integral XRB.
 
Curves in Fig.~\ref{r5_r7} show the loci
of the thermal emission: the upper curve indicates colours
of the bremsstrahlung, while the lower curve -- plasma emission with
metal abundances reduced to $20$\,\% of cosmic abundances.
Labels give temperatures in keV. The X-ray colours of the enhancements
found by \cite{soltan02} in the RASS maps 
are shown with open symbols. The open square denotes the colours
around the Abell clusters (the dotted ellipse represents the
$90$\,\% confidence limits), and open circles -- around galaxies.
The enhancements are integrated
between $0\fdg3$ and $2\fdg1$ for clusters, while for the Lick counts
the results are shown for $4$ separation bins: $<0\fdg3$, $0\fdg3 - 0\fdg7$,
$0\fdg7 - 1\fdg5$, and $1\fdg5 - 3\fdg1$, where the smaller symbol in
Fig.~\ref{r5_r7} refers to the larger separation. For both
samples $1\degr$ corresponds to several Mpc and it is likely that some
contribution to the X-ray halos detected in the RASS is generated not only
by the diffuse WHIM emission, but also by the hot gas in poor groups of galaxies
which surround the rich Abell clusters and have typical temperatures
of the order of $10^7$\,K. 

Stars in the upper left corner show the colours of $25$
``normal'' clusters of galaxies selected from the \ROSAT\ archives.
Clear separation of clusters and the extended emission in Fig.~\ref{r5_r7}
demonstrates that at least substantial fraction of
the diffuse emission correlated with galaxies does not originate in the
``unknown'' clusters but represents a distinct component of the soft XRB.

\begin{figure}
\resizebox{\hsize}{!}{\includegraphics[width=0.8\linewidth]{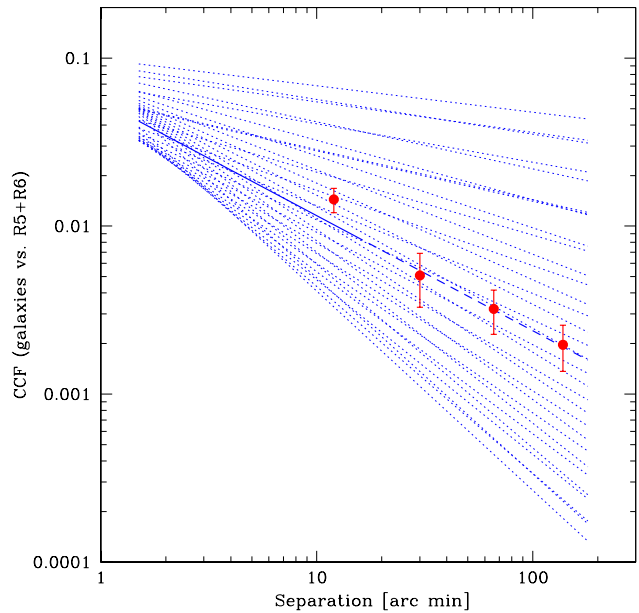}}
\caption{The cross-correlation function of the \ROSAT\ PSPC bands R5 + R6 and
the galaxy sample: the best fit - thick line, selection of power law fits
allowed by $68$\,\% confidence limits - thin lines; plots for separations
greater than $15\arcmin$ are extrapolated. Points with the error bars show
the CCF of the RASS maps (R5 + R6) with the Lick counts of galaxies (see the
text for details).
\label{pspc_ccf_2}
}
\end{figure}

Next, we have tried to fit the observed variations of the 
the X-ray surface brightness enhancements around  galaxies by a power law:
\begin{equation}
\rho(\theta) = a\, \theta^\gamma + \rho_{o}\,,
\end{equation}
with three a priori unknown parameters $a$, $\gamma$ and $\rho_o$, where
$\rho_o$ represents the XRB component uncorrelated with galaxies. 
In this case, however, the signal-to-noise ratio was too low to obtain
restrictive limits for the interesting parameters. The best fit values for the
slope $\gamma$ were equal to $-0.7$ for bands R5 and R6 and $-1.3$ for R7,
but the range of uncertainties was wide at any reasonable significance level.
To improve statistics, we have repeated all the calculations using merged
bands R5 and R6. The allowed range of fits still covers a large area in parameter
space (Fig.~\ref{pspc_ccf_2}). This is mostly because the data cover a narrow range
of separations. Additionally, we do not have a priori sufficiently accurate
information on $\rho_o$ and all three parameters have to be fitted
simultaneously. 

In Fig.~\ref{pspc_ccf_2} we plotted also the CCF of the RASS with the 
galaxy distribution of the Lick counts (see \cite{soltan02} for
the details of calculations). Error bars represent $1\sigma$ uncertainties
based on the simulations. Near perfect agreement between the \ROSAT\
pointings and the All-Sky Survey
is to some extent accidental, since the galaxy samples in both investigations
have different statistical properties; in particular they have different
magnitude limits and undoubtedly different redshift distributions.
It is evident, however, that estimates of the CCF at small separations
are consistent with the extrapolation based on the RASS.

\subsection{EPIC/MOS}

Search for the WHIM emission in the EPIC/MOS observations was realized
in a similar way as with the PSPC data. First, we have determined the
$\rho(\theta)$ in four energy bands. Te eliminate residual effects introduced
by vignetting, we have compared the real data with data from
simulations. In a single simulation run the distribution of galaxies
in each pointing was randomized and the calculations of $\rho(\theta)$
were performed in the same way as for the real data. Then, the CCF of
the X-ray and galaxy distributions was obtained by subtraction of the averaged
simulated countrates $\rho(\theta)$ from the observed data. For four energy
bands 30 full simulation runs have been performed. The rms scatter between
simulated functions $\rho(\theta)$ was used as estimator of uncertainties
in the real data.

\begin{figure}
\resizebox{\hsize}{!}{\includegraphics[width=0.8\linewidth]{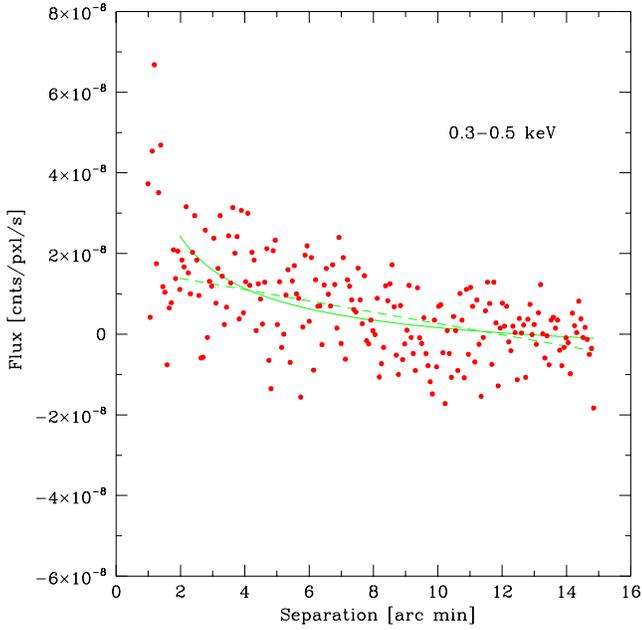}}
\caption{The cross-correlation function of the galaxy sample and the EPIC/MOS
data in the energy band $0.3-0.5$\,keV for separations $1\arcmin - 15\arcmin$.
The dashed line shows the linear fit to the data points between $2\arcmin$ and
$15\arcmin$, dashed curve -- the power law fit with $\gamma=-1$. 
\label{xmmccf_1}
}
\end{figure}

\begin{table}
\caption[ ]{Excess countrates}
\vspace*{0.10cm}
\begin{tabular}{ccc}
\hline \hline
 Band & Energy (keV) & Countrates [cnt/pxl/s]          \\\hline
 X1   & 0.3 -- 0.5   & $ (1.80 \pm 0.23)\times 10^{-8} $ \\   
 X2   & 0.5 -- 1.0   & $ (1.50 \pm 0.30)\times 10^{-8} $ \\  
 X3   & 1.0 -- 2.0   & $ (0.37 \pm 0.42)\times 10^{-8} $ \\
 X4   & 2.0 -- 4.5   & $ (0.62 \pm 0.38)\times 10^{-8} $ \\
\hline
\end{tabular}
\label{fluxes}
\end{table}

\begin{figure}
\resizebox{\hsize}{!}{\includegraphics[width=0.8\linewidth]{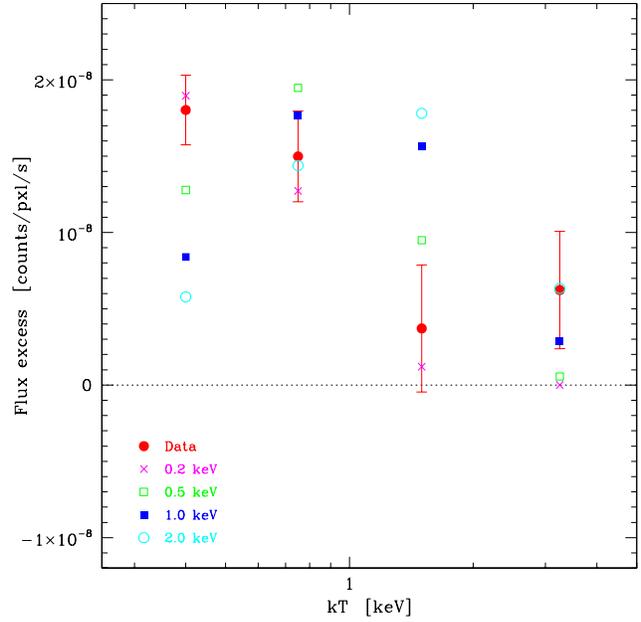}}
\caption{Full circles with $1\sigma$ error bars -- average excess surface
brightness measured $2\arcmin$ from the galaxy in the sample in 4 energy
bands based on EPIC/MOS data; remaining symbols -- expected countrates
of thermal bremsstrahlung for $0.2$, $0.5$, $1.0$ 1nd $2.0$\,keV with
normalization fitted to the observational points.
\label{temperatures}
}
\end{figure}

Despite the large number of pointings used in the present calculations,
the signal-to-noise ratio of the correlation measurement is still low. In
Fig.~\ref{xmmccf_1} the CCF of the energy band X1 ($0.3-0.5$\,keV) is shown
with the two best fits to the data points for separations between
$2\arcmin$ and $15\arcmin$: linear (dashed line) and power law with
$\gamma = -1$ (solid curve). Clearly, the S/N ratio is too low to
distinguish between those two solutions. Nevertheless, the CCF signal
is detected at a decent significance level. The slope of the linear
fit differs from zero by more than $7\sigma$.

Using the slope determination we estimated the average excess countrate
at separation of $2\arcmin$ above the level measured at $15\arcmin$.
The results for four energy bands are listed in the Table~\ref{fluxes} and
shown in Fig.~\ref{temperatures}. Then, we have convolved the thermal
bremsstrahlung
spectra with the effective area of the X-ray telescope/EPIC MOS detectors
system and fitted the resulting counts in four energy bands to the observed
signal. The predicted counts of thermal emission for ${\rm k}T = 0.2$,
$0.5$, $1.0$ and $2.0$\,keV are also plotted in Fig.~\ref{temperatures}.

The spectrum of the observed enhancements correlated with the galaxy sample
is roughly consistent with the thermal emission with ${\rm k}T \sim 0.2$\,keV
and becomes incompatible with ${\rm k}T \ga 0.5$\,keV. The amplitude of
the signal is, however, very weak. In the band X1 ($0.3-0.5$\,keV) it
amounts roughly just to $1.3$\,\% of the total countrate. 

Although, both the PSPC and EPIC/MOS data show traces of the WHIM emission,
the amplitude of the signal obtained from the PSPC is roughly by a factor of 2
greater than from the EPIC/MOS. We have compared the amplitudes of the correlated
emission in the band R5 and X2, as they cover similar energy ranges.
Assuming a temperature of the bremsstrahlung emission ${\rm k}T = 0.5$\,keV,
and correcting for effects of the absorption by cold gas in the
Galaxy\footnote{The average hydrogen column density in the \ROSAT\ sample
amounts to $2.25\times 10^{20}$\,cm$^{-2}$, while in the XMM sample --
$2.77\times 10^{20}$\,cm$^{-2}$.} the excess surface brightness in the
energy band $0.5-1.0$\,keV measured by \ROSAT\ is equal to
$(1.41\pm 0.26)\times 10^{-13}$\,erg/(cm$^2 \cdot$s$\cdot$deg$^2$),
while the XMM-Newton data provide
$(0.71\pm 0.14)\times 10^{-13}$\,erg/(cm$^2 \cdot$s$\cdot$deg$^2$).
Two factors could contribute to this difference. First, the observational
material is heterogeneous. It spans a wide range of exposure times.
In particular, the XMM-Newton sample covers
more than $17$\,deg$^2$, but half of the counts comes from just above
$3.6$\,deg$^2$. The distribution of
exposures in the \ROSAT\ sample is also very nonuniform. Consequently,
the effective area of the survey is much smaller than the total area
and it is possible that the effect of cosmic variance is partially
responsible for different results.

Second, the galaxies selected from NED also constitute highly heterogeneous
sample. Although the selection criteria for the PSPC and
EPIC/MOS observations were identical, the average apparent magnitudes
and surface density of galaxies are substantially different in both
samples. The average galaxy magnitude selected in the
EPIC/MOS data is fainter by $0.4$\,mag than in the PSPC data. It is
difficult to assess impact of those systematic effects, but it is
likely that galaxies selected in the EPIC/MOS observations are on the
average more distant than in the \ROSAT\ observations and the measured
CCF signal at fixed angular separation is weaker.

The amplitudes measured by both instruments differ by $2.4\sigma$, where
$\sigma$ denotes just the statistical uncertainty introduced by both
instruments. To compare measurements reduced to physical units one needs
assessments of systematic errors which in the present investigation
are not well determined.

\section{WHIM emission - prospects for the quantitative study \label{future}}

Extensive search for fluctuations of the soft XRB using two excellent
instruments -- \ROSAT\ PSPC and XMM-Newton EPIC/MOS -- has led to the
detection of extremely weak signal which most likely represents the WHIM
emission. Although virtually all the available observational material
(several hundreds of pointings) has been used, the signal-to-noise
ratio does not allow for the quantitative study of physical properties
of the WHIM. A strong dependence of the observed signal on energy
is highly suggestive and -- if the thermal mechanism is assumed -- indicates
temperatures below $0.5$\,keV. Still, the data are insufficient to determine
the density or temperature distributions of the emitting gas. 

Correlation analysis provides information on the average amplitude and
temperature of the WHIM emission. Simulations show, however, that both
density and temperature of the intergalactic medium cover a wide range
of magnitudes (\citealt{bryan}, \citealt{croft}).
To investigate physical parameters of the WHIM, one needs
sensitive observations of a definite WHIM cloud associated with the
individual concentration of galaxies. The present analysis shows
that in the soft XRB the WHIM signal amounts to $\sim 1$\,\% of the
total XRB. It seems that such a low threshold for the diffuse emission is
below the sensitivity of the present-day instruments using wide energy
bands. However, presence of strong lines in the thermal bremsstrahlung
of optically thin plasma at temperatures expected for the intergalactic
plasma could make the investigation of the WHIM possible. To test the feasibility
of this approach we plan to search for characteristic spectral features
of the XRB in the vicinity of selected galaxies using the EPIC/MOS
observations.

\vspace{2mm}
ACKNOWLEDGMENTS. 
The \ROSAT\ project has been supported by the 
Bundesministerium f\"ur Bildung, Wissenschaft, Forschung und Technologie
(BMBF/DARA) and by the Max-Planck-Gesellschaft (MPG). We thank all people
involved in the XMM-Newton project for making the XMM Science Archive and
Standard Analysis System friendly environments.
AMS thanks the Max-Planck-Gesellschaft for support.
This research has made use of the NASA/IPAC Extragalactic Database (NED) which
is operated by the Jet Propulsion Laboratory, California Institute of
Technology, under contract with the National Aeronautics and Space
Administration.
This work has been partially supported by the Polish KBN grant 1~P03D~003~27.

\end{document}